\documentclass[aps,prl,twocolumn,amsmath,amssymb,showpacs,showkeys]{revtex4}
\usepackage{epsfig}
\usepackage{epstopdf}
\usepackage{prettyref}
\usepackage{graphicx}
\usepackage{color}
\definecolor{red}{rgb}{1,0,0}
\begin {document}
\title {Effect of Distributed Particle Magnetic Moments on the Magnetization of NiO Nanoparticles}
\author{S. D. Tiwari}
\affiliation{School of Physics and Materials Science, Thapar University, Patiala 147004, India}
\author{K. P. Rajeev}
\affiliation{Department of Physics, Indian Institute of Technology Kanpur 208016, India}
\begin{abstract}
Magnetization of nanoparticles of NiO are measured and analyzed taking into account a distribution in particle magnetic moment. We find that disregarding this  distribution in the analysis is the reason for the many anomalous observations reported on this system in the literature.
\end{abstract}
\pacs{75.50.Ee, 75.50.Tt, 75.75.-c, 75.75.Fk}
\keywords{nanoparticle, superparamagnetism, magnetization, susceptibility}
\maketitle
Studies on particles of a few nanometers in size have been attracting
quite a lot of interest of late. For a physicist the attraction is
mainly due to the emergent properties that a particle or a collection of
them shows when the particle size is made very small. According to
N\'{e}el tiny particles of an antiferromagnetic material should exhibit
magnetic properties such as superparamagnetism and weak ferromagnetism
\cite{Low Temp. Phys.}. If the surface to volume ratio, which varies as
the reciprocal of particle size, of an antiferromagnetic particle
is made sufficiently large then it can have a nonzero net magnetic
moment because of an imperfect cancellation of elementary moments
pointing in different directions near the surface of the particle.
Recently nanoparticles of antiferromagnetic materials have gained quite
a lot of attention mainly because they show some surprising and unusual
behavior unobservable in ferro or ferrimagnetic nanoparticles
\cite{Kodama 1997, Morup 2004}. Among different antiferromagnetic
nanoparticles, NiO is a comparatively more interesting and well studied
system \cite {Kodama 1997, Richardson and Milligan, Richardson 1991,
Makhlouf 1997}. Long back, Richardson and Milligan \cite { Richardson
and Milligan} reported magnetization measurements on NiO nanoparticles
of different sizes and this has been followed by many more reports
\cite{Kodama 1997, Richardson 1991, Makhlouf 1997}, mostly by other
workers. We also reported some work on NiO nanoparticles \cite{SDT 2005,
SDT 2006} where we found that, at low temperatures, this system shows
spin glass behavior. We also showed that the low temperature behavior of
NiO nanoparticles is not superparamagnetic contrary to popular belief
and expectation. After our work some other workers have also reached similar
conclusions on NiO nanoparticles independently \cite{Winkler, Vijay}. 

The magnetization of antiferromagnetic nanoparticles is expected to be 
described by a modified Langevin function \cite{Makhlouf 1997, Kilcoyne, 
Makhlouf, Seehra 2000}. However, fitting the magnetization data of bare 
antiferromagnetic NiO nanoparticles as a function of magnetic field to 
the modified Langevin function results in unphysical fit parameters. 
For instance, the estimated particle magnetic moment turns out to be 
about 2000~$\mu_{\textnormal{B}}$ for 5.3~nm particles \cite {Makhlouf 1997}. Such values 
for the particle magnetic moments are much larger than the expected value of about a few hundred Bohr magnetons from uncompensated spins on the surface of particles. Another method for the estimation of 
particle magnetic moment of NiO nanoparticles has also been used 
\cite{Kodama 1997}. Here the value of saturation magnetization at low 
temperature is divided by the estimated  number of particles to get the 
particle magnetic moment. However the moment estimated 
by this method has also been found to be larger than the expected value. A multisublattice model has been proposed to explain this large value of 
particle magnetic moment for NiO nanoparticles though there is no experimental support for this model yet \cite{Kodama 1997}. 

The magnetization of nanoparticles of magnetic materials is expected to be 
only a function of the applied magnetic field $B$ and temperature $T$ 
and should scale with $(\frac{B}{T})$ above the bifurcation temperature ($T_{\textnormal{bf}}$) between low field cooled and zero field cooled magnetization. The magnetization of NiO 
nanoparticles is not found to show this scaling \cite{Makhlouf 1997}. 
These observations motivated us to revisit the antiferromagnetic NiO 
nanoparticles system once again. This work is an attempt to find the 
reason for getting unphysical numbers for the particle magnetic moment of 
NiO nanoparticles when the traditional methods are used to analyze the 
magnetization data above $T_{\textnormal{bf}}$. The value of the bifurcation temperature $T_{\textnormal{bf}}$ for the present system is about 295~K in 100~G applied magnetic field \cite{SDT 2006}.

Here we present the magnetization measurements on 5~nm bare NiO
particles as a function of applied magnetic field at
sufficiently high temperatures, but well below
the N\'{e}el temperature ($T_{\textnormal{N}}$) of the system which is known to be about 
523~K \cite{Smart}. We analyzed the data using the modified
Langevin function  without considering any distribution in
particle magnetic moment and then repeated the analysis taking into account a distribution in
the particle moment. We were pleasantly surprised by the results as it turned out that we can account for the anomalous observations reported on this system by the earlier workers.

NiO nanoparticles were prepared by a sol-gel method by reacting in
aqueous solution nickel nitrate and sodium hydroxide at pH = 12 at
room temperature as described elsewhere \cite{Richardson and
Milligan, Makhlouf 1997, Richardson 1991}. We used nickel (II)
nitrate hexahydrate   (99.999\%), sodium hydroxide pellets
(99.99\%), both from Aldrich, and triple distilled water to make
nickel hydroxide. The sample of nickel oxide nanoparticles was
prepared by heating the nickel hydroxide at 523~K for 3 hours in
flowing helium gas (99.995\%). The sample was characterized by
x-ray diffraction and transmission electron microscopy. The
average crystallite size as well as the particle size were found to be
about 5~nm. The details of sample synthesis and structural
characterization have been reported in one of our earlier works \cite{SDT
2006}. All the magnetic measurements were done with a commercial SQUID
magnetometer (Quantum Design, MPMS XL5).
\begin{figure}[thb]
\begin{center}
\includegraphics[angle=0,width=1.\columnwidth]{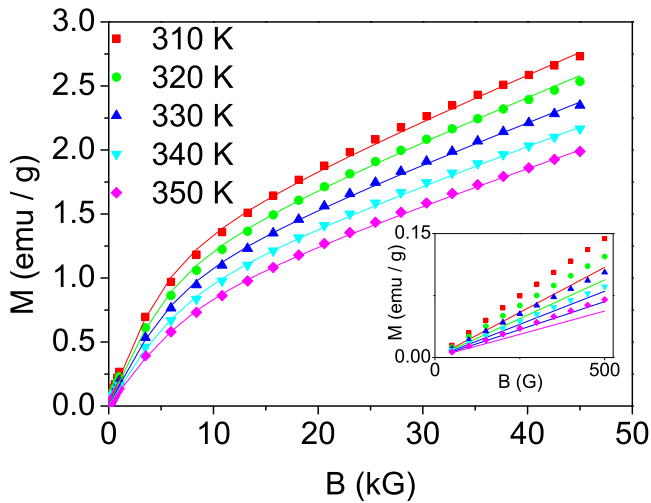}
\caption{(Color online) Magnetization as a function of applied magnetic field for 5~nm NiO particles at different temperatures. Solid lines show fits to Equation(\ref{eq:lang-modi}). The inset shows a magnified view of the data along with the fits at lower magnetic fields which makes it clear that the fit quality is  poor there.}
\label{fig:mh-5nm}
\end{center}
\end{figure}

We measured the magnetization $M$ of the 5~nm NiO particles as a
function of external applied magnetic field $B$ at different
temperatures $T$ ($T_{\textnormal{bf}} < T < T_{\textnormal{N}}$). These measurements are shown in Figure
\ref{fig:mh-5nm}. Small particles of magnetic materials
are expected to be superparamagnetic at sufficiently high temperature ($T > T_{\textnormal{bf}}$)\cite{Low Temp. Phys.}.
Magnetization $M$ of a superparamagnetic material as a function of magnetic field $B$ and temperature $T$ is described by \cite{Bean}
\begin{equation}
\label{eq:lang}
M = M_0 L(x),
\end{equation}
where $L(x) = [\coth{(x)} - (\frac{1}{x})]$ is the Langevin
function and $x = \frac{\mu_{\textnormal{p}} B}{k_{\textnormal{B}} T}$. Here $M_0$ is the
saturation magnetization, $\mu_{\textnormal{p}}$ is the particle magnetic moment
and $k_{\textnormal{B}}$ is the Boltzmann constant. However, magnetization of small
particles of antiferromagnetic materials, even though they are 
superparamagnetic  cannot be described by
Equation~(\ref{eq:lang}). Rather, they are well described by an
altered form known as the
modified Langevin function \cite{Kilcoyne, Makhlouf,Seehra 2000},
which contains an extra linear term in $B$ i.e.
\begin{equation}
\label{eq:lang-modi}
M = M_0 L(x) + \chi_{\textnormal{a}} B,
\end{equation}
where $\chi_{\textnormal{a}}$ is the susceptibility of randomly oriented
antiferromagnetic particle cores. We fitted the data shown in Figure~\ref{fig:mh-5nm} to the above equation and the resulting fits are shown as solid lines in the figure. Values of fit
parameters $M_{0}$, $\mu_{\textnormal{p}}$ and $\chi_{\textnormal{a}}$  obtained are
presented in Table~\ref{table-310-350K-5nm}. 
\begin{table}[htb]
\caption{Values of fit parameters $M_{0}$, $\mu_{\textnormal{p}}$ and
$\chi_{\textnormal{a}}$ to Equation (\ref{eq:lang-modi}) and the corresponding values of the goodness of fit parameter  
$R^{2}$ for the 5~nm NiO nanoparticles at different temperatures.}
\label{table-310-350K-5nm} \vspace{0.2in}
\begin{tabular}{|c|c|c|c|c|}
\hline
$T$&$M_{\textnormal{0}}$&$\mu_{\textnormal{p}}$&$\chi_{\textnormal{a}}$ ($10^{-6}$&$R^{2}$\\
(K)&(emu/g)&($\mu_{\textnormal{B}}$)&emu/g Oe)&\\
\hline
310&1.30&1967&34&0.9993\\
320&1.18&1841&32&0.9994\\
330&1.05&1825&31&0.9995\\
340&0.94&1734&29&0.9996\\
350&0.82&1621&27&0.9997\\
\hline
\end{tabular}
\end{table}

From Figure \ref{fig:mh-5nm} we see that the fits 
to Equation (\ref{eq:lang-modi}) more or less pass
through the measured data points which means the fit quality is quite 
good. A statistical measure of the goodness
of a fit is the coefficient of determination $R^{2}$. The closer
$R^2$ is to unity the better the fit. Values of the coefficient of
determination $R^{2}$ for the fits 
are shown in Table \ref{table-310-350K-5nm} along with the fit 
parameters. The $R^2$ values are greater than 0.999 in all cases and this  
once again confirms the good quality of the fits.

From Table
\ref{table-310-350K-5nm} we see that the particle magnetic moment
$\mu_{\textnormal{p}}$ is about two thousand Bohr  magnetons and it decreases
with increasing temperature. Somewhat similar results have been
reported by others on NiO nanoparticles \cite{Makhlouf 1997} as well as
on NiO nanorods \cite{Seehra 2004}. However there is a glaring inconsistency between
the numbers we get for $M_0$ and $\mu_{\textnormal{p}}$.  In
Equation (\ref{eq:lang-modi}) the fit parameters $M_0$ and $\mu_{\textnormal{p}}$ should be
related as $M_0 = N \mu_{\textnormal{p}}$  where $N$ is the number of particles per unit mass
of the sample. If we use this relation to estimate the average particle
magnetic moment $\mu_{\textnormal{p}}$ it turns out to be about 60~$\mu_{\textnormal{B}}$ at 320~K which
is very small compared to the value of $\mu_{\textnormal{p}}$ presented in Table
\ref{table-310-350K-5nm}. This fact invalidates the fit of experimental data to Equation~(\ref{eq:lang-modi}). In the next paragraph we shall give another argument against the numbers presented in Table~\ref{table-310-350K-5nm}.

N\'{e}el had discussed various ways by which a magnetic moment can appear on an NiO particles due to incomplete compensation of atomic moments in different sublattices \cite{Low Temp. Phys., Richardson 1991}. According to  N\'{e}el the
particle magnetic moment of an NiO nanoparticle can be written as
\begin{equation}
\label{eq:neil-moment}
\mu_{\textnormal{p}} = p~\mu_{\textnormal{A}}~\mu_{\textnormal{B}}.
\end{equation}
Here $p$ is a number  which depends on the size, crystal structure and form or shape of the particle and $\mu_{\textnormal{A}}$ is the magnetic moment of Ni$^{2+}$ ion which is known to be 3.2~$\mu_{\textnormal{B}}$ \cite{Kittel}.
N\'{e}el proposed a total of five different possibilities for the
origin of magnetic moment on an NiO particle of approximately cubic shape. Out of the five, two
possibilities give the values of particle magnetic moment to be
zero corresponding to $p=0$. The other three
possible values of $p$ are about $n^{\frac{1}{3}}$, $n^{\frac{1}{2}}$
and $n^{\frac{2}{3}}$ where $n$ is the number of Ni$^{2+}$ ions in the
particle. Now, there can be many other possible set of values of $p$ corresponding to shapes such as spheres, tetrahedrons or, more realistically, irregular shapes. But it seems safe to assume that the upper limit of $p$ will be about $n^{\frac{2}{3}}$ and the average will be considerably less.
Making use of  the fact that the crystal structure of NiO is
face centered cubic with lattice constant 4.176~$\AA$, the value
of $n$ for a 5~nm diameter spherical particle turns out to be about 3592. Using the information presented above along with Equation~(\ref{eq:neil-moment})
we get the particle magnetic moment to be about 49~$\mu_{\textnormal{B}}$,
191~$\mu_{\textnormal{B}}$ and 750~$\mu_{\textnormal{B}}$ corresponding to $p$ values of
$n^{\frac{1}{3}}$, $n^{\frac{1}{2}}$ and $n^{\frac{2}{3}}$
respectively. 
Thus according to N\'{e}el's picture the maximum possible
value of particle magnetic moment for  a 5~nm NiO particle would be about
750~$\mu_{\textnormal{B}}$ which is much smaller than the values shown in Table \ref{table-310-350K-5nm}.

While using Equation (\ref{eq:lang-modi}) to fit the magnetization
data we made a tacit assumption that all the particles have the same magnetic
moment. But, this is not true. A sample of NiO
nanoparticles has a distribution of particle magnetic moments not
only due to distribution in size and shape but also due to the way in which the imbalance of spins arises in the particle as pointed out by N\'{e}el. We hazard the guess that disregarding the distribution in particle moments is perhaps the reason for the unphysical fit parameters obtained in Table
\ref{table-310-350K-5nm}. Now we would like to carry out this fitting taking into account a moment distribution. Making our path easier is a precedent in this kind of analysis set by Silva et. al. in analyzing the magnetization of ferritin, a biological antiferromagnetic nanoparticle system \cite{Silva}.

In a sample that has a distribution in particle magnetic
moments the low field magnetization is governed by the particles
with larger magnetic moments. The contribution of particles of lower
magnetic moments to the magnetization becomes important only at higher applied
fields where the high field forces the moments to align with the field. The
effect of a distribution in the particle magnetic moment on the
magnetization of the system will show up in the Langevin or modified Langevin fit. To see this dependence let us take a look at the comparison of the measured magnetization to the modified Langevin fit at low fields  in the inset of Figure
\ref{fig:mh-5nm}. It is clear that the fits are no good at 
low fields while from the main panel one can see that the situation is much better
at higher fields. This low field misfit is an indication of the role
a distribution in the particle magnetic moment plays on the
magnetization of a system. In this case the modified Langevin function has clearly underestimated the contribution of the larger moments.

Following Silva et. al. we assume that in a  sample of nanoparticles the distribution in particle magnetic
moment, $\mu$, can be described by a log normal distribution function of the form \cite{Silva}
\begin{equation}
\label{eq:lognormal-distribution}
f(\mu) = \frac{1}{\mu s \sqrt{2 \pi}} \exp -\frac{[{\ln (\frac{\mu}{n})]}^2}{2 s^2},
\end{equation}
where $n$ and $s$ are parameters that characterize the distribution.
The mean particle magnetic moment $\mu_{\textnormal{mean}}$ is equal to $n
\sqrt{e^{s^2}}$. Now, Equation (\ref{eq:lang-modi}) takes on the new form
\begin{equation}
\label{eq:lang-modi-lognormal} M (B, T) = N \int_{0}^{\infty} \mu
L(x) f(\mu) d\mu + \chi_a B,
\end{equation}
where $N$ is the total number of particles contributing to
magnetization and $L(x)$ is the Langevin function with $x = \frac{\mu B}{k_{\textnormal{B}} T}$. We used this
equation to fit the magnetization data of Figure
\ref{fig:mh-5nm} and the fit parameters thus obtained along
with the values of coefficient of determination $R^2$ are shown in
Table \ref{table-310-350K-5nm-lognormal}.
\begin{table}
\caption{Values of fit parameters $N$, $s$, $n$ and $\chi_{a}$ to
Equation (\ref{eq:lang-modi-lognormal}) and the values of $R^{2}$
for the 5~nm NiO particles at different temperatures. The mean magnetic moment of a particle estimated from the fit parameters are also shown.}
\label{table-310-350K-5nm-lognormal} \vspace{0.2in}
\begin{tabular}{|c|c|c|c|c|c|c|}
\hline
$T$&$N$ ($10^{17}$&$s$&$n$&$\chi_a (10^{-6}$&    $\mu_{\textnormal{mean}}$     &$R^{2}$\\
(K)&/g)&&$(\mu_\textnormal{B})$&emu/g~Oe)&                 $(\mu_\textnormal{B})$       &\\
\hline
310&8.7&1.34&116.5&21.4&287&0.999997\\
320&7.3&1.30&130.3&21.4&305& 0.999998\\
330&6.5&1.27&136.8&20.9&308&0.999997\\
340&6.9&1.28&118.9&19.8&271&0.999998\\
350&6.8&1.26&114.4&18.9&254&0.999998\\
\hline
\end{tabular}
\end{table}

\begin{figure}[thb]
\begin{center}
\includegraphics[angle=0,width=1.0\columnwidth]{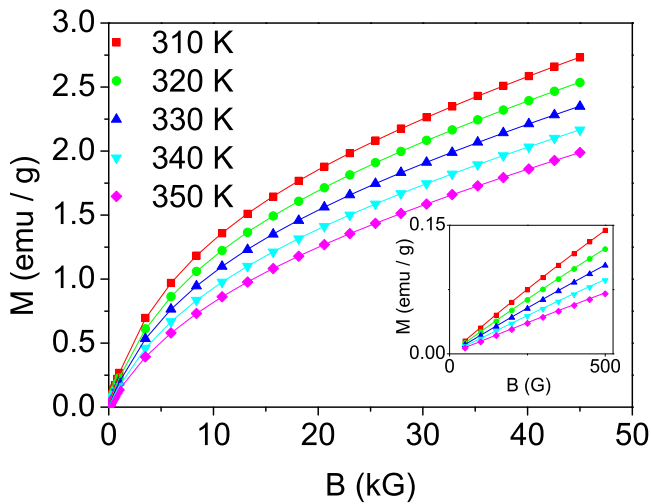}
\caption{(Color online) Magnetization as a function of applied magnetic field for 5~nm NiO particles at different temperatures. Solid lines show the fits of data to Equation~(\ref{eq:lang-modi-lognormal}). The inset shows the fits at low magnetic fields.}
\label{fig:mh-5nm-modi-lang-fit-distri}
\end{center}
\end{figure}

We see in Table \ref{table-310-350K-5nm-lognormal} that the
values of the coefficient of determination $R^2$ are surprisingly
greater than 0.99999 in all the cases. These values of $R^2$ are much better than the values of the same shown in Table
\ref{table-310-350K-5nm}. This means that the quality of the fits
using Equation (\ref{eq:lang-modi-lognormal}) is much better than
that using Equation (\ref{eq:lang-modi}). In Figure
\ref{fig:mh-5nm-modi-lang-fit-distri} we show the new fits which clearly look good.  The inset clearly shows that the
fits are very good for lower values of applied magnetic fields
also where the effect of distribution in the particle magnetic
moment on the magnetization of the system is important as already
discussed. We can now conclude that the fits of the magnetization data to Equation
(\ref{eq:lang-modi-lognormal}) are much better than that to
Equation (\ref{eq:lang-modi}).

From Table \ref{table-310-350K-5nm-lognormal} we find that the mean particle magnetic moments, which are equal to $n\sqrt{e^{s^2}}$, turns out to be about a few hundred Bohr magnetons, rather than a few thousand Bohr magnetons as found in the earlier fit, and are also well below the maximum possible value of about 750~$\mu_{\textnormal{B}}$ according to the models proposed by N\'{e}el for the origin of magnetic moment in NiO particles.

In conclusion, we reported magnetization as a function of applied
magnetic field for 5~nm NiO particles at different temperatures above
the bifurcation temperature $T_{\textnormal{bf}}$. Fitting the magnetization data to the modified
Langevin function without considering any distribution in the particle
magnetic moments yields very large and unphysical values for the
particle magnetic moment. However we got reasonable values for the particle
magnetic moment if the modified Langevin function is used to fit the
magnetization data taking into account a distribution in the particle
magnetic moment. The distribution in the particle magnetic moment arises
from a distribution in particle size and form. 
This work clearly shows that the non consideration of
a distribution in particle magnetic moment could be the reason for the
anomalously large values of magnetic moment of NiO nanoparticles
reported in the literature.

\end{document}